\documentclass[aip,jmp,preprint,showpacs]{revtex4-1}
\usepackage{graphicx,dcolumn,array,bm,amsmath,amssymb}

\newcommand{\nc}{\newcommand}               %
\nc{\nuc}[2]	{$^{#1}${#2}}		    
\nc{\hatf}[3]   {\hat{{#1}_{#2}}(#3)}       
\nc{\hatfpm}[4] {\hat{{#1}}_{#2}^{(#4)}(#3)}
\nc{\fcpm}[4]   {{#1}_{#2}^{C(#4)}(#3)}     
\nc{\csop}      {U(\theta)}                 
\nc{\csopd}     {U^{\dagger}(\theta)}       
\nc{\bra}	{\langle}		    
\nc{\ket}	{\rangle}		    
\nc{\br}	{{\bf r}}		    %
\nc{\bp}	{{\bf p}}		    %
\nc{\bk}	{{\bf k}}		    %
\nc{\bhr}	{{\bf{\hat r}}}		    %
\nc{\bhp}	{{\bf{\hat p}}}		    %
\nc{\bR}	{{\bf R}}		    %
\nc{\lam}	{\lambda}		    %
\nc{\al}	{\alpha}		    %
\nc{\gam}	{\gamma}	            %
\nc{\vc}[1]	{\bf #1}	            %
\nc{\Nc}	{M}	                    %
\nc{\NcB}	{M^{B}}	                    %

\begin{document}

\title{Coulomb-distorted plane wave: partial wave expansion and asymptotic forms}

\author{I. Hornyak}
\email{hornyak.istvan@atomki.mta.hu}
\affiliation{University of Debrecen, Faculty of Informatics, PO Box 12, 
4010 Debrecen, Hungary}

\author{A.T. Kruppa}
\email{kruppa.andras@atomki.mta.hu}
\affiliation{Institute of Nuclear Research, Bem t\'er 18/c, 
4026 Debrecen, Hungary}

\date{\today}

\begin{abstract}
Partial wave expansion of the Coulomb-distorted plane wave is determined and studied. 
Dominant and sub-dominant  asymptotic expansion 
terms are given and leading order three-dimensional asymptotic form is derived.
The generalized hypergeometric function $_2F_2(a,a;a+l+1,a-l;z)$ is expressed 
with the help of confluent hypergeometric functions and the asymptotic expansion  of $_2F_2(a,a;a+l+1,a-l;z)$
is simplified.
\end{abstract}

\maketitle

\section{Introduction\label{intro}}

Scattering of charged particles is an important and  difficult topic in quantum mechanics. 
Even the two-body case is exceptional
since it does not fit into the conventional time-dependent scattering theory approach. 
Several methods are proposed to overcome 
the difficulties caused
by the long range behavior of the Coulomb interaction. 
In the time-dependent framework the formalism of Dollard \cite{dol64} modifies the Moller wave operator.  In stationary approach the method of van Haaringen\cite{har76} 
introduced Coulomb asymptotic states
and Mulherin and Zinnes \cite{mul70} promoted distorted asymptotic states. The connection of the latter 
two formalism was also studied \cite{bar89}. 
Following references \cite{Kad05,Kad09} we will call the asymptotic state of Mulherin and Zinnes
as Coulomb-distorted plane wave (CDPW). 

The recently developed surface integral formalism of quantum scattering theory \cite{Kad05,Kad09,bra12} 
is valid not only for two-body scattering but it 
can be extended to scattering of three charged particles.  Another advantage of this formalism is that it handles 
the short and longe range interactions on an
equal footing. It is said in \cite{Kad09} ``All the results ... rely on the asymptotic forms of the plane
wave and the Coulomb-distorted plane waves". The CDPW indeed  
plays an important role in the surface integral formalism. In this framework 
the full scattering wave function is composed of as a sum of the CDPW and a so called scattered part.
Furthermore the surface integral expression of the scattering amplitude
also refers to the CDPW.


A three-dimensional leading order asymptotic form of the CDPW 
was given in \cite{Kad05}. Mathematical interpretation of this result was presented
in reference \cite{Kad09} in section 5. The asymptotic expansion can be interpreted 
in distributional sense and the properties of the considered test function space plays an 
important role. Based on our convergent partial wave (pw) expansion
of the CDPW and the asymptotic form of the pw components we will 
derive a three-dimensional leading order asymptotic form for the CDPW. Our expression  
is in agreement 
with the result of Ref. \cite{Kad05,Kad09} if the test function
space given by Taylor \cite{tay74} is used.

Interestingly the pw expansion of the CDPW has been given only recently \cite{hor12}. 
In this work the pw components are applied for the description of scattering process of two particles based on the
complex scaling method.
Here we 
further study the pw decomposition and derive four different analytical forms. The asymptotic expansion 
of the pw component given in \cite{hor12}
contains a recursion relation for the expansion coefficients. In this paper we derive an explicit expression for them.

Our results concerning the CDPW are presented at Chapter \ref{cdpw}.
Since the generalized
hypergeometric function  $_2F_2(a,a;a+l+1,a-l;z)$ plays a substantial role in the formalism 
therefore in the Appendix 
the properties of this function are examined in detail.
The summary is given at chapter \ref{conc}.

\section{Partial wave expansion and asymptotic forms\label{cdpw}}

We write the pw expansion of the Coulomb-distorted plane wave in the form
\begin{equation}\label{taupw}
e^{i\bk\br}(kr \mp\bk\br)^{\pm i\gamma}=\sum_{l=0}^\infty (2l+1)\tau^{(\pm)}_l(\gamma,kr)P_l(\cos(\vartheta)),
\end{equation}
where the upper and lower signs correspond to the post and prior form of the CDPW.
The angle between the vectors $\bk$ and $\br$ is signed by $\vartheta$ and the Legendre polynomial
is denoted by $P_l(x)$. 
The real variable  $\gamma$ in (\ref{taupw}) is the Sommerfeld parameter and  $\bk$ is the momentum vector. The bold face letters $\bk$ and $\br$ denote elements of the 
three-dimensional Euclidean space $\mathbb{R}^3$, while $k$ and $r$ denote the corresponding magnitudes.
The radial function $\tau^{(\pm)}_l(\gamma,kr)$ can be calculated by the integral
\begin{equation}\label{tau}
\tau^{(\pm)}_l(\gamma,kr)=\frac{1}{2}(kr)^{\pm i\gamma}\int_{-1}^1e^{\pm ikrx} (1\mp x)^{\pm i\gamma} P_l(x) dx.
\end{equation}
A compact expression for the pw component $\tau^{(\pm)}_l(\gamma,kr)$ can be given for arbitrary $l$. 
Using (\ref{tau}) and the formula 2.17.5.6 in \cite{prud} we get 
\begin{equation}\label{taudefpost}
\tau^{(+)}_l(\gamma,kr)=\frac{(-i\gamma)_l}{(1+i\gamma)_{l+1}}
(2kr)^{i\gamma}e^{ikr}\ _2F_2(1+i\gamma,1+i\gamma;l+2+i\gamma,1+i\gamma-l;-2ikr)
\end{equation}
and
\begin{equation}\label{taudefprior}
\tau^{(-)}_l(\gamma,kr)=(-1)^l \frac{(i\gamma)_l}{(1-i\gamma)_{l+1}}
(2kr)^{-i\gamma}e^{-ikr}\ _2F_2(1-i\gamma,1-i\gamma;l+2-i\gamma,1-i\gamma-l;2ikr),
\end{equation}
where $(a)_n$ is the Pochhammer symbol and $_2F_2$ is the generalized hypergeometric function\cite{abra}. 
From the the explicit forms (\ref{taudefpost}) and (\ref{taudefprior}) one can notice that 
$\tau^{(-)}_l(\gamma,kr)=(-1)^l\tau^{(+)}_l(\gamma,kr)^*$. 
In order to further study the function $\tau^{(\pm)}_l(\gamma,kr)$  the properties of the function 
$_2F_2(a,a;a+l+1,a-l;z)$ are investigated in the Appendix.
In the case of the post form of CDPW the following identifications can be made $a=1+i\gamma$ and
$z=-2ikr$ and in the prior form we have $a=1-i\gamma$, $z=2ikr$. Using Theorem 1 of the Appendix 
and the formula (\ref{taudefpost}) we can write down three
equivalent expressions for the partial wave component
 \begin{eqnarray}\label{tauful1}
& \tau^{(+)}_l(\gamma,kr)=\frac{e^{ikr+\gamma\pi/2}}{2ikr} \sum_{n=0}^{l} (-1)^n \binom{l}{n} \frac{(l+1)_n}{n!}\; 
\gamma(1+i\gamma+n,2ikr) (2ikr)^{-n}\;,\\
& \tau^{(+)}_l(\gamma,kr)=(-1)^l (2kr)^{i \gamma} e^{ikr} \sum_{n=0}^{l} (-1)^n 
\binom{l}{n} \frac{(l+1)_n}{(1+i\gamma)_{n+1}}\; {}_1 \textnormal{F}_1 (1+i\gamma,2+i\gamma+n;-2ikr)\;
 \end{eqnarray}
and 
 \begin{eqnarray}\label{tauful2}
& \tau^{(+)}_l(\gamma,kr)= \frac{e^{\gamma\pi/2}}{2ikr}\left(
e^{ikr}\kappa_l^+(1+i\gamma,-2ikr) + e^{-ikr} \kappa_l^-(1+i\gamma,-2ikr)\right).
 \end{eqnarray}
 
In charge-less case (i.e. $\gamma=0$) the CDPW goes into an ordinary plane wave. Using Propositions 2 of the Appendix we 
get the expected result for the pw components i.e. $\tau^{(\pm)}_l(0,kr)=i^lj_l(kr)$ i.e. the post and
prior forms are identical.

Using Theorem 2 of the Appendix we get for the asymptotic expansion of 
$\tau^{(\pm)}_l(\gamma,kr)$ as $r\rightarrow\infty$ 
\begin{equation}\label{tauasyfulp}
 \begin{aligned}
 \tau^{(+)}_{l}(\gamma,kr)\sim &\frac{e^{ikr}}{2ikr}e^{\gamma\pi/2}\ \Gamma(1+i\gamma)
 \ _3F_1\left (1+i\gamma,-l,l+1;1;(2ikr)^{-1}\right )\\
&- \frac{e^{-ikr}}{2ikr}(2kr)^{i\gamma}(-1)^l\sum_{n=0}^\infty d_n^{(l)} \frac{2^n}{(-2ikr)^n}\;.
\end{aligned}
\end{equation}
and 
\begin{equation}\label{tauasyfulm}
 \begin{aligned}
 \tau^{(-)}_{l}(\gamma,kr)\sim &-(-1)^l\frac{e^{-ikr}}{2ikr}e^{\gamma\pi/2}\ \Gamma(1-i\gamma)
\  _3F_1\left (1-i\gamma,-l,l+1;1;(-2ikr)^{-1}\right )\\
&+\frac{e^{ikr}}{2ikr}(2kr)^{-i\gamma}\sum_{n=0}^\infty \left (d_n^{(l)}\right )^* \frac{2^n}{(2ikr)^n}\;.
\end{aligned}
\end{equation}
The expansion coefficients $d_n^{(l)}$ satisfy the recursion (\ref{rec}-\ref{rec1}) and 
the solution of the recursion is given by (\ref{recsol}). 
In these equations we have to identify $a$ by $1+i\gamma$ ($1-i\gamma$) for the post (prior) form of the pw 
component. The expressions (\ref{tauasyfulp}) and (\ref{tauasyfulm}) show both the dominant and sub-dominant 
terms and the nature of the terms are determined by the 
sign of ${\rm Im}(ikr)$. 
Instead of the study of the explicit form of $\tau^{(\pm)}_{l}(\gamma,kr)$ 
the asymptotic expansions (\ref{tauasyfulp}) and (\ref{tauasyfulm}) can be obtained 
directly from (\ref{tau}) with a straightforward but tedious calculation 
based on the theorem 227 of Ref. \cite{fou}.

Now we derive a three-dimensional leading order asymptotic form for the CDPW.
Substituting expressions (\ref{tauasyfulp}) and (\ref{tauasyfulm}) into (\ref{taupw}) and keeping only the leading order
terms we get our result in the form of a distribution 
\begin{equation}\label{asy3d}
e^{i\bk\br}(kr \mp\bk\br)^{\pm i\gamma}\sim \pm\frac{2\pi}{ikr}\left (e^{\pm ikr}e^{\gamma\pi/2}\Gamma(1\pm i\gamma)
\delta(\hat\br\mp\hat\bk)
- e^{\mp ikr}(2kr)^{\pm i\gamma}\delta(\hat\br\pm \hat\bk)\right).
\end{equation}
It is known that the prior form of the CDPW can be simply obtained from the post form expression: one has to 
replace $\bk$ with $-\bk$ and make complex conjugation. According to (\ref{asy3d}) this procedure is 
valid also for the asymptotic forms.

In charge-less case the three-dimensional asymptotic expression (\ref{asy3d}) goes into the well known form \cite{mes}
\begin{equation}\label{asy3dexp}
e^{i\bk\br}\sim \frac{2\pi}{ikr}\left (e^{ikr}\delta(\hat\br-\hat\bk)
- e^{-ikr}\delta(\hat\br+\hat\bk)\right).
\end{equation}
On the test function spaces $D^{\pm}$ of reference \cite{Kad09} our result goes into equations (197) and (198) of \cite{Kad09}
\begin{equation}\label{asy3dd}
e^{i\bk\br}(kr \mp\bk\br)^{\pm i\gamma}\sim 
\mp\frac{2\pi}{ikr}e^{\mp ikr}(2kr)^{\pm i\gamma}\delta(\hat\br\pm\hat\bk)\textnormal{ on }D^\pm.
\end{equation}
On test functions belong to $D^\pm$ the contribution of the first term of (\ref{asy3d}) is zero.
Our rigorously derived result (\ref{asy3d}) differs from the expression of Ref. \cite{Kad05,Kad09}.
However the surface integral formalism, as explained in reference \cite{Kad09}, depends only on 
the validity of (\ref{asy3dd}). 

\section{Summary}\label{conc}
The generalized hypergeometric function $_2F_2(a,a;a+l+1,a-l;z)$ is expressed as a finite sum of confluent hypergeometric
functions. Three equivalent expressions are determined and so the numerical evaluation can be simplified. Dominant and
sub-dominant terms are identified in the asymptotic expansion. A closed expression is given for the coefficients of the asymptotic expansion.
Using the previous results four equivalent expressions are determined for the partial wave component of the Coulomb-distorted
plane wave. The asymptotic expansion of the partial wave component is deduced. Rigorously derived leading order three-dimensional
asymptotic form of the Coulomb-distorted plane wave is given.
\begin{acknowledgments}
The publication was supported by the T\'AMOP-4.2.2.C-11/1/KONV-2012-0001 project. 
The project has been supported by the European Union, co-financed by 
the European Social Fund.
\end{acknowledgments}

\appendix*
\section{The hypergeometric function $_2F_2(a,a;a+l+1,a-l;z)$\label{2f2chapter}}

We will express the function $_2F_2(a,a;a+l+1,a-l;z)$ in terms of finite linear combination of confluent hypergeometric functions.
In order to do this we need a proposition.\\
{\bf 1. Proposition.}  Let $l$ be a non-negative integer and let $\mathbb{X}_l:=\{ x\in\mathbb{Z} \;\arrowvert\; x\leq l \}$. If 
$z\in\mathbb{C}\backslash\{0\}$ and  $a\in\mathbb{C}\backslash\mathbb{X}_l$ then
 \begin{equation}\label{prep2}
 \sum_{k=0}^{l} (-1)^k \binom{l}{k} (l+1)_k (1-a)_k \; {}_1 \textnormal{F}_1 (-k,a-k;z) \frac{z^{-k}}{k!}=
     (-1)^l\ _3\textnormal{F}_1(1-a,-l,l+1;1;-1/z) .
 \end{equation}
{\bf Proof.} The definition of the confluent hypergeometric function can be turned into the form
 \[
 {}_1 \textnormal{F}_1 (-k,a-k;z) = \sum_{s=0}^k (-1)^s \binom{k}{s} \frac{z^s}{(a-k)_s}\;.
 \]
Using this expression we get for the the left hand side of (\ref{prep2})
\[
 \sum_{k=0}^l\sum_{s=0}^k (-1)^{k+s} \binom{l}{k}\binom{k}{s} \frac{(l+1)_k (1-a)_k}{k! (a-k)_s} z^{s-k}\;.
 \]
We can rearrange the summation indexes and get
 \begin{equation} \label{prep2a}
 (-1)^l \sum_{s=0}^{l} \binom{l}{s} (l+1)_s (1-a)_s \frac{z^{-s}}{s!}\; \varepsilon_l(s)\;,
 \end{equation}
where
\[
 \varepsilon_l(s)=\sum_{k=0}^{l-s} (-1)^k \binom{l-s}{k} \frac{(l+1)_{l-k} s!}{(l+1)_s (l-k)!}.
 \]
Using the formula 0.160.2 of \cite{grad} we get $\varepsilon_l(s)=1$
and we can recognize that the summation in (\ref{prep2a}) gives the function $\ _3\textnormal{F}_1(1-a,-l,l+1;1;-1/z)$.\hspace*{\fill}$\Box$\\
{\bf 1. Theorem.}  Let $l$ be a non-negative integer and let $z\in\mathbb{C}\backslash\{0\}$,  
$a\in\mathbb{C}\backslash\mathbb{Z}$ then the function $_2F_2(a,a;a+l+1,a-l;z)$ can be given in the following forms
\begin{eqnarray}
&\label{theo1a}(-1)^l \frac{(a)_{l+1}}{(1-a)_l} \sum_{k=0}^{l} (-1)^k \binom{l}{k} \frac{(l+1)_k}{(a)_{k+1}}\;
{}_1 \textnormal{F}_1 (a,a+k+1;z)\; ,\label{eq1}\\
&\label{theo1b}(-z)^{-a} \frac{(a)_{l+1}}{(1-a)_l} \sum_{k=0}^{l}  \binom{l}{k} 
\frac{(l+1)_k}{k!}\;\gamma(a+k,-z)\; (z)^{-k}\; ,\\
&\label{theo1c}\frac{(a)_{l+1}}{(1-a)_l}(-z)^{-a}\left(\kappa^+_l(a,z) +e^z\kappa^-_l(a,z)\right)\;.
\end{eqnarray}
Here we introduced the notations 
\begin{equation}\label{kappam}
 \kappa^-_l(a,z)=(-1)^{l+1}\sum^l_{n=0}\frac{(l+n)!}{n!(l-n)!}
 \frac{(-1)^n}{z^n}U(-a+1,-a-n+1;-z)
 \end{equation}
and
\begin{equation}\label{kappap}
\kappa^+_l(a,z)=\Gamma(a)\ _3F_1(a,-l,l+1;1;-1/z).
\end{equation}
The confluent hypergeometric function of the second kind \cite{abra} and  
the incomplete gamma function \cite{abra} are signed by $U(a,b,z)$ and $\gamma(a,z)$, respectively.
\\
{\bf Proof.} First we show that (\ref{theo1a}) equals (\ref{theo1b}). 
Using the identity 5.3.5.4 of \cite{prud} and the expression 7.11.3.1 of \cite{prud} we get for (\ref{eq1})
 \[
 (-z)^{-a} \frac{(a)_{l+1}}{(1-a)_l} \sum_{k=0}^l\sum_{s=0}^k (-1)^{l+k+s} \binom{l}{k}\binom{k}{s} \frac{(l+1)_k}{k!} \;\gamma(a+s,-z) (-z)^{-s}\;.
 \]
Rearranging the summation we have 
 \[
 (-z)^{-a} \frac{(a)_{l+1}}{(1-a)_l} \sum_{s=0}^{l} (-1)^s \binom{l}{s} \frac{(l+1)_s}{s!}\;\gamma(a+s,-z)\; (-z)^{-s}\;\varepsilon_l(s)
 \]
and using that $\varepsilon_l(s)=1$ we get (\ref{theo1b}). 

Next we show that  $_2F_2(a,a;a+l+1,a-l;z)$ equals  (\ref{theo1b}).
Using the 
\begin{equation}
 {}_p F_q
 \left(\left.
 \begin{array}{cc}
 (a_{p-1}), & \sigma +l\\
 (b_{q-1}), & \sigma
 \end{array}
 \right| z \right) = 
 \sum_{k=0}^{l} \binom{l}{k} \frac{z^k}{(\sigma)_k} \frac{\prod (a_{p-1})_k}{\prod (b_{q-1})_k}
 {}_{p-1} F_{q-1}
 \left(\left.
 \begin{array}{c}
 (a_{p-1})+k\\
 (b_{q-1})+k
 \end{array}
 \right| z \right)
\end{equation}
identity, which was proved in \cite{kar70}, we get for $_2F_2(a,a;a+l+1,a-l;z)$
 \begin{equation}\label{2f2a}
 (-z)^{-a} \sum_{k=0}^l\sum_{s=0}^l (-1)^{k+s} \binom{l}{k}\binom{l}{s} \frac{(a)_{l+1}}{l! (a-l)_k}\; \gamma(a+k+s,-z) (-z)^{-s}\;.
 \end{equation}
The following expression, valid for $k>0$, can be easily established using the formula 8.356.9 of \cite{grad}
 \begin{equation}\label{grec}
 \gamma(b+k,z)=(b)_k\gamma(b,z)-e^{-z}z^{b}\sum_{m=0}^{k-1}\frac{(b)_k}{(b)_{m+1}}z^m\;
 \end{equation}
and for $k=0$ we have $\gamma(b+0,z)=(b)_0\gamma(b,z)=\gamma(b,z)$. Choosing $b=a+s$ in (\ref{grec}) and substituting it into 
(\ref{2f2a}) we get 
\begin{equation}
 \begin{aligned}
 &(-z)^{-a} \frac{(a)_{l+1}}{(1-a)_l} \sum_{s=0}^{l} (-1)^s \binom{l}{s} \frac{(l+1)_s}{s!}\;\gamma(a+s,-z)\; (-z)^{-s}+\\
 &- \frac{(a)_{l+1}}{l!}\; e^{z}\sum_{k=1}^l\sum_{m=0}^{k-1} (-1)^k \binom{l}{k} \frac{\Gamma(a+k)\Gamma(l+m+1-k)(-z)^m}{(a-l)_k \Gamma(l+a+m+1)} 
 \frac1{\Gamma(-[k-1-m])}
 .\label{2f2b}
\end{aligned}
 \end{equation}
In the derivation of this expression we used 0.160.2 of \cite{grad}. We have to consider the case  $a\in\mathbb{C}\backslash\mathbb{Z}$ and 
$l\geq k\geq 1$ so we have that $k-1-m$ is a
non-negative integer.  We get that the double sum in (\ref{2f2b}) is zero and this means that $_2F_2(a,a;a+l+1,a-l;z)$ 
is identical with (\ref{theo1b}). 

Finally we show that (\ref{theo1c}) equals (\ref{theo1a}).
Applying the expression 7.2.2.2 of \cite{prud} we get for (\ref{theo1c}) 
\begin{equation}
\frac{(a)_{l+1}}{(1-a)_l}\left (\sum_{n=0}^{l} (-1)^{n+l} \binom{l}{n} \frac{(l+1)_n}{(a)_{n+1}}\; 
{}_1 \textnormal{F}_1 (a,a+1+n;z)+(-z)^{-a}\Gamma(a)\epsilon_l (a,z)\right ) \;,
 \end{equation}
where
\begin{alignat*}{2} 
 \epsilon_l (a,z) =& _3F_1(a,-l,l+1;1;-1/z)\\
                           &  -(-1)^l \sum_{n=0}^{l} (-1)^n \binom{l}{n} (l+1)_n (a)_n \; {}_1 \textnormal{F}_1 (-n,-a+1-n;z) 
			   \frac{z^{-n}}{n!} \;.
 \end{alignat*}
According to Proposition 1 $\epsilon_l (a,z)=0$ and so we derived (\ref{theo1a}) from (\ref{theo1c}).\hspace*{\fill}$\Box$

In order to show that the pw component of CDPW is proportional to the spherical Bessel function in the case of charge-less particles we 
will need the following statement.\\
{\bf 2. Proposition.}  Let $l$ be a non-negative integer and let $z\in\mathbb{C}\backslash\{0\}$ and 
$a\in\mathbb{C}\backslash\mathbb{Z}$ then
\begin{equation}
\lim_{a\to 1}\ (1-a)_l\ _2F_2(a,a;a+l+1,a-l;z)=i^l(l+1)!\;e^{z/2}j_l(iz/2).
\end{equation}
{\bf Proof.} It is easy to establish that $(1)_{l+1}e^z\kappa_l^-(1,z)/z=-i^l(l+1)!e^{z/2}h_l^{(2)}(iz/2)/2$ and 
$(1)_{l+1}\kappa_l^+(1,z)/z=-i^l(l+1)!e^{z/2}h_l^{(1)}(iz/2)/2$ using 10.1.16 and 10.1.17 of \cite{abra}. From these facts 
and from (\ref{theo1c}) the claim follows.\hspace*{\fill}$\Box$
\\
{\bf 2. Theorem.} Let $l\ge 0$ integer, $a\in \mathbb{C}$, $0 < {\rm Re}(a)<l+2$ 
then the asymptotic expansion of the function $_2F_2(a,a;a+l+1,a-l;z)$ at $\vert z\vert\rightarrow\infty$ 
can be written in the form
\begin{equation}\label{sajatasy}
\begin{aligned}
&_2F_2(a,a;a+l+1,a-l;z)\sim\frac{(a)_{l+1}}{(1-a)_l}(-z)^{-a}\kappa^+_l(a,z)+\\
&(-1)^{l}\frac{(a)_{l+1}}{(1-a)_l}\frac{e^z}{z}\sum_{n=0}^\infty 
 \frac{(1-a)_{n}(a)_l}{(a-n)_l}\ {}_3F_2
 \left(\left.
 \begin{array}{ccc}
-l, & -l, & -n\\
1, & 1-a-l
\end{array}
\right| 1 \right)\frac{1}{z^n}.
\end{aligned}
\end{equation}
{\bf Proof.}
We will use the expression (\ref{theo1c}) to determine the asymptotic 
expansion. First we notice that  $\kappa_l^+(a,z)$
is a finite sum, it is a polynomial of order $l$ in the variable $1/z$. This means that  
$\frac{(a)_{l+1}}{(1-a)_l}(-z)^{-a}\kappa_l^+(a,z)$ can be considered  as an asymptotic expansion 
at $\vert z\vert\rightarrow\infty$ with respect to the asymptotic sequence \cite{erd}  
$z^{-a}/z^n$. 
The function $\kappa_l^-(a,z)$, according to equation (\ref{kappam}), is a finite linear combination  of  
confluent hypergeometric functions of second kind and since the asymptotic expansion of $U(a,b,z)$ is given by 15.5.2 of
\cite{abra} we can properly rearrange the summation indexes and we can write 
\begin{equation}\label{asy1}
\kappa_l^-(a,z)\sim(-1)^{l}\frac{(-z)^a}{z}\sum_{n=0}^\infty 
\left(\sum_{m=0}^{\min(n,l)}(-1)^{m}\frac{(l+m)!}{m!(l-m)!}\frac{(m+1)_{n-m}(1-a)_{n-m}}{(n-m)!}\right) \frac{1}{z^n}.
\end{equation}
Using the formulas 7.2.3.15 and 7.3.5 of \cite{prud} we can carry out the summation over $m$ 
on the right hand side of (\ref{asy1}) and get 
\begin{equation}\label{asy2}
\kappa_l^-(a,z)\sim(-1)^{l}\frac{(-z)^a}{z}\sum_{n=0}^\infty 
\frac{(1-a)_{n}(a)_l}{(a-n)_l}\ {}_3F_2
 \left(\left.
 \begin{array}{ccc}
-l, & -l, & -n\\
1, & 1-a-l
\end{array}
\right| 1 \right)\frac{1}{z^n}.
\end{equation}
\hspace*{\fill}$\Box$\\
The expressions (\ref{sajatasy}) shows that if ${\rm Im}(z)>0$ than the dominant term is given
by the second term of the right hand side of (\ref{sajatasy})
and if ${\rm Im}(z)<0$ then 
the roles of the terms in (\ref{sajatasy}) interchange.  

The complete asymptotic expansion of the function $_2F_2(a_1,a_2;b_1,b_2;z)$ 
was given by Luke  in chapter 5.11.3 of reference \cite{luke}. However in our case 
the numerator parameters are equal to each others and we can not use this result. 
We rewrite the function $_2F_2(a,a;a+l+1,a-l;z)$ in terms of Meijer's G function (see 5.11.1(1) in \cite{luke})
\begin{equation}
_2F_2(a,a;a+l+1,a-l;z)=\frac{(-1)^l(a)_{l+1}}{(1-a)_l}
G_{2,3}^{1,2}\left(-z \left| 
\begin{array}{ccc}
1-a, & 1-a\\
0, & -a-l, & l+1-a
\end{array}
\right.\right).
\end{equation}
For the asymptotic expansion of the Meijer's G function we can use formula 5.10(10) of \cite{luke} and 
if we apply 5.7.(13-15) and 5.9.2 of Ref.\cite{luke} we get 
\begin{equation}\label{2f2asy}
_2F_2(a,a;a+l+1,a-l;z)\sim -\frac{(-1)^l(a)_{l+1}}{(1-a)_l} H_{2,2}(-z).
\end{equation}
Taking into account 5.11.1(18) and 5.11.3(4) in \cite{luke} we can write
\begin{equation}\label{h22}
H_{2,2}(-z)=-\frac{e^{z}}{z}\sum_{n=0}^\infty  d^{(l)}_{n} 2^n\frac{1}{z^n}\;.
\end{equation}
The coefficients $d^{(l)}_{n}$ satisfy the recurrence relation (see 5.11.3(6) in \cite{luke})
\begin{equation}\label{rec}
 4(n+1)d^{(l)}_{n+1}=2(2n^2-n(2a-3)-l^2-l-a+1)d^{(l)}_{n}-n(n-l-a)(n+l+1-a)d^{(l)}_{n-1} 
\end{equation}
with initial conditions (see 5.11.1(18-20) in \cite{luke})
\begin{equation}\label{rec1}
d^{(l)}_{0}=1\;,\ \ \ \  d^{(l)}_{1}=\frac{1}{2}\left(1-l^2-l-a\right).
\end{equation}
The asymptotic expansion (\ref{2f2asy}) is valid in the region 
$-\frac{\pi}{2}+\delta\le{\rm arg}(-z)\le\frac{3\pi}{2}-\delta$ where $\delta>0$. 

We will show that our asymptotic expansion (\ref{sajatasy}) and formula (\ref{2f2asy}) are identical in the considered $z$ region.
Among the two terms at the right hand side of (\ref{sajatasy}) 
the second one is the dominant in the considered $z$
region and the first one can be neglected.
To rewrite (\ref{2f2asy}) and (\ref{h22}) we have to solve the recurrence relation (\ref{rec}).
\\
{\bf 3. Theorem.} Let $l$ and $n$ be non-negative integers and let $a\in\mathbb{C}$, $0<{\rm Re}\left(a\right)<l+2$ then 
the solution of the recurrence relation (\ref{rec}) with initial conditions (\ref{rec1}) is
 \begin{equation}\label{recsol}
 d^{(l)}_{n} =  \frac{(1-a)_{n}(a)_l}{2^n(a-n)_l}\ {}_3F_2
 \left(\left.
 \begin{array}{ccc}
-l, & -l, & -n\\
1, & -a-l+1
\end{array}
\right| 1 \right).
 \end{equation}
{\bf Proof.} We prove this theorem by induction. 
With direct calculation it is easy to check that the claim is satisfied for $n=0,1,2$. Assuming that (\ref{recsol}) is true for 
$n=s$ and $n=s+1$ the claim, after some trivial simplification, is reduced to the validity of the equation 
\begin{align*}
 (s+2)&(s+2-l-a) {}_3F_2\left(\left.\begin{array}{ccc}-l, & -l, & -s-2\\1, & -a-l+1\end{array}\right| 1 \right)=\\
    =&\left[ 2(s+1)^2-(s+1)(2a-3)-l^2-l-a \right] {}_3F_2\left(\left.\begin{array}{ccc}-l, & -l, & -s-1\\1, & -a-l+1\end{array}\right| 1 \right)\\
	&-(s+1)(s+2+l-a){}_3F_2\left(\left.\begin{array}{ccc}-l, & -l, & -s\\1, & -a-l+1\end{array}\right| 1 \right)\;.
 \end{align*}
This equation however is known to be true \cite{wolfram}. \hspace*{\fill}$\Box$

Using Theorem 3 and equation (\ref{h22}) we see that (\ref{2f2asy}) and the second term on the right hand side of 
(\ref{sajatasy}) are indeed identical.

\end{document}